\documentclass{ifacconf}

\usepackage{graphicx}      
\usepackage{natbib}        
\usepackage{url}
\usepackage[scr=boondoxo,scrscaled=1.05]{mathalfa}
\usepackage{amsmath}
\usepackage{amsfonts}
\usepackage{amssymb}
\usepackage{upgreek}
\usepackage{bm}

\let\classAND\AND
\let\AND\relax
\usepackage{algorithmic}

\let\AND\classAND
\AtBeginEnvironment{algorithmic}{\let\AND\algoAND}

\usepackage{algorithm}
\usepackage{xcolor}
\usepackage{booktabs}

\begin{document}
\begin{frontmatter}

\title{Nodal Hydraulic Head Estimation through Unscented Kalman Filter for Data-driven Leak Localization in Water  Networks\thanksref{footnoteinfo}} 

\thanks[footnoteinfo]{The authors would like to thank the Spanish national project Project L-BEST under Grant PID2020-115905RB-C21 funded by MCIN/ AEI /10.13039/501100011033.
This work was also supported by a grant of the Romanian Ministry of Research, Innovation and Digitization, CCCDI - UEFISCDI, project number PN-III-P2-2.1-PED-2021-1626, within PNCDI III. This work has been submitted to IFAC for possible publication.} 

\author[First]{Luis Romero-Ben} 
\author[UB,ILDS]{Paul Irofti} 
\author[Third]{Florin Stoican}
\author[First,Fourth]{Vicenç Puig}

\address[First]{Institut de Robòtica i Informàtica Industrial, CSIC-UPC, Llorens i Artigas 4-6, 08028, Barcelona, Spain \\(e-mail: luis.romero.ben@upc.edu).}
\address[UB]{LOS-CS-FMI,
University of Bucharest, Romania \\(e-mail: paul@irofti.net)}
\address[ILDS]{Institute for Logic and Data Science, Bucharest, Romania}
\address[Third]{Department of Automation Control and Systems Engineering, Politehnica University of Bucharest, Romania, \\(e-mail: florin.stoican@upb.ro)}
\address[Fourth]{Supervision, Safety and Automatic Control Research Center (CS2AC) - UPC, Campus de Terrassa, Terrassa, 08222, Barcelona, Spain, (e-mail: vicenc.puig@upc.edu)}

\begin{abstract}                
In this paper, we present a nodal hydraulic head estimation methodology for water distribution networks (WDN) based on an Unscented Kalman Filter (UKF) scheme with application to leak localization. The UKF refines an initial estimation of the hydraulic state by considering the prediction model, as well as available pressure and demand measurements. To this end, it provides customized prediction and data assimilation steps. Additionally, the method is enhanced  by dynamically updating the prediction function weight matrices. Performance testing on the Modena benchmark under realistic conditions demonstrates the method's effectiveness in enhancing state estimation and data-driven leak localization.
\end{abstract}


\end{frontmatter}

\section{Introduction}


The appearance of leaks in water distribution networks (WDN) results in a significant water loss of approximately 126 billion cubic meters per year
worldwide (expressed as non-revenue water), as indicated by \cite{Liemberger2019}. For decades, water utilities have harnessed software-based techniques for leak detection and localization. Within the approaches that analyse the steady-state of the WDN, three main categories emerge: model-based, data-driven and mixed model-based/data-driven methods.

Model-based methods exploit a hydraulic model to emulate the behaviour of the WDN, calibrating both network characteristics and nodal demands. The aim is to obtain simulated hydraulic data that can be compared with actual measurements from the real network \citep{ Sophocleous2019}. 
Mixed model-based/data-driven methods reduce the dependence on the model through machine learning algorithms while maintaining the node-level accuracy \citep{Capelo2021}. 
Finally, data-driven approaches remove the necessity of a hydraulic model, 
using algorithms that exploit sensor data to provide leak location areas. Within data-driven schemes, a well-established family of methods relies on interpolating the complete hydraulic state from available measurements and topology, to then perform leak localization through the reconstructed schemes. State estimation has been successfully utilized in WDNs with techniques such as Kriging interpolation \citep{Soldevila2020}, graph neural networks \citep{Xing2022} and graph signal processing \citep{Zhou2023}.  

This article introduces the next step of the interpolation methodology presented in \cite{RomeroBen2022b}, known as Graph-based State Interpolation (GSI), and its subsequent evolution introduced in \cite{Irofti2023}, known as Analytical Weighting GSI (AW-GSI). 
In particular, a nodal hydraulic head estimation methodology is proposed based on an Unscented Kalman Filter (UKF) scheme with application to leak localization. The proposed methodology uses the UKF scheme to improve the head estimation from an input reconstructed state vector, leading to various improvements compared to the previous GSI schemes: 

\textit{The available pressure and demand information is fused}. Previous state estimation methods often rely on pressure data due to the reduced cost and ease of installation of pressure meters. The continuous upgrade of the urban infrastructures leads to the rise of "smart cities", characterized by advanced real-time metering capabilities \citep{Ali2022}.  Automated Metering Readers (AMR) play a crucial role by measuring real-time consumption. Thus, a state estimation strategy that neglects the processing of demand data risks the loss of critical information. 

\textit{The graph diffusion weights from the prediction function are dynamically updated}. These weights are not only configured considering the WDN structure and the physics governing its behaviour, but they are also adapted to the current estimated state. This drastically improves the state estimation performance in comparison to previous methodologies.

The structure of the paper is as follows: In Section \ref{sec:preliminaries}, the methods that will be used to develop the proposed approach are presented. In Section \ref{sec:approach}, this approach is introduced. A case study is described in Section \ref{sec:case_study} accompanied by simulations in Section \ref{sec:results}. Finally, Section \ref{sec:conclusions} draws the main conclusions and suggests future research paths. 


\section{Preliminaries} \label{sec:preliminaries}

Let us 
model the network topology 
by a graph $\mathcal{G}=(\mathcal{V},\mathcal{E})$, where $\mathcal{V}$ is the node set (reservoirs and junctions), and $\mathcal{E}$ denotes the edge set (pipes). An arbitrary node is represented as $\mathscr{v}_i\in\mathcal{V}$, whereas an arbitrary edge is denoted as $\mathscr{e}_k = (\mathscr{v}_i,\mathscr{v}_j)\in\mathcal{E}$. The latter represents the link between nodes $\mathscr{v}_i$ and $\mathscr{v}_j$, with $\mathscr{v}_i$ as its source and $\mathscr{v}_j$ as its sink. The nodal hydraulic heads are selected as representatives of the network states, gathered in $\bm h$.

\subsection{State interpolation} An initial attempt to estimate the hydraulic state of a WDN was proposed in \cite{RomeroBen2022b} through GSI, which only requires hydraulic head data and structural information. This method estimates the complete network state by solving the optimization problem 
\begin{subequations} \label{eq:1}
\begin{align}
\label{eq:1_a}\min_{\bm{h}} \quad & \frac{1}{2}\big[\bm{h}^T\bm{L_d}\bm{h}+\alpha \gamma ^2\big]\\
\textrm{s.t.} \quad & \bm{\hat{B}}\bm{h}\leq\boldsymbol{\upgamma}, \ \; \ 
\gamma > 0, \ \; \ 
\bm{S}\bm{h}=\bm{h_s},  
\end{align}
\end{subequations}
where $\bm{L_d}$  is a Laplacian-based matrix generated from the WDN underlying graph, $\bm{\hat{B}}$ is an approximated incidence matrix of $\mathcal{G}$, $\gamma$ is a slack variable and $\bm{S}$ and $\bm{h_s}$ are the sensorization matrix and the head measurement vector respectively. 
Although the details are presented in \cite{RomeroBen2022b}, let us highlight that $\bm{L_d}=(\bm{D}-\bm{W})\bm{D}^{-2}(\bm{D}-\bm{W})$, where $\bm{W}$ is the weighted adjacency matrix of $\mathcal{G}$, with $w_{ij} = 1/\rho_{k}$ if $\mathscr{e}_k = (\mathscr{v}_i,\mathscr{v}_j)\in\mathcal{E}$ and $w_{ij} =0$ otherwise, and $\rho_{k}$ is the length of the pipe represented by edge $\mathscr{e}_k$. $\bm{D}$ is the degree matrix, a diagonal array  obtained as $d_{ii} = \sum_{j=1}^{n} w_{ij}$.
Briefly, GSI pursues the closest state vector to fulfill that $\bm{h} = \bm{D}^{-1}\bm{W}\bm{h}$, where $\bm{D}^{-1}\bm{W}\bm{h}$ diffuses the state considering the structure of $\mathcal{G}$, while contemplating directionality and measurements-related constraints.

Recently, a novel physical-based weighting process was designed to improve GSI, leading to a new interpolation method that is referred to as AW-GSI. This weighting process is based on the linearization of the Hazen-Williams equation \citep{Sanz2016}, and yields a new weighted adjacency matrix $\bm{W^{AW}}$ as follows: 
\begin{equation}\label{eq:3}
    w^{AW}_{ij}(\tilde{h}_i,\tilde{h}_j) = \sigma_{k}^{0.54}\left[m_{kj}(\tilde{h}_i-\tilde{h}_j)\right]^{-0.46}, 
\end{equation}
where $\sigma_k=(\mu_{k}^{1.852} \delta_{k}^{4.87})/(10.67 \rho_{k})$ is the pipe conductivity (in S.I.) for $\mathscr{e}_k$, with $\mu_{k}$ and $\delta_{k}$ being the roughness and diameter of the pipe respectively. Note that $\bm{W^{AW}}$ is richer in both structural and hydraulic information in comparison to $\bm{W}$, hence leading to an improvement in the accuracy of the state estimation. Besides, $m_{kj}$ is the $k\mbox{-}j$ element of the incidence matrix $\bm{M}\in\mathbb{R}^{|\mathcal{E}|\times n}$, defined as:
    \begin{equation}\label{eq:8}    m_{kj}=\begin{cases}\hphantom{-}1,& \tilde{h}_i\geq  \tilde{h}_j \;\; (\mathscr{e}_k = (\mathscr{v}_i,\mathscr{v}_j)\in \mathcal{E});\\ -1,& \tilde{h}_i< \tilde{h}_j \;\; (\mathscr{e}_k = (\mathscr{v}_j,\mathscr{v}_i)\in \mathcal{E});\\ \hphantom{-}0,& \textrm{if $\mathscr{v}_i$ and $\mathscr{v}_j$ are not adjacent}\end{cases}
    \end{equation}
The characteristics of $\bm{\tilde{h}}$ are introduced in \textit{Lemma 1} at \cite{Irofti2023}. Due to the ultimate goal of performing leak localization, we selected $\tilde{h}_i=h^{nom}_i$ and $\tilde{h}_j=h^{nom}_j$, with $\bm{h}^{nom}$ being the leak-free reference, as most localization techniques operate over pressure residuals, i.e., difference of pressure between leak and leak-free scenarios. 
This selection of $\tilde{\bm{h}}$ leads to the following quadratic programming problem  
\begin{align}
\label{eq:2}\min_{\bm{\Delta h}} \quad & \frac{1}{2}\big[\bm{\Delta h}^T\bm{L_d^{AW}}\bm{\Delta h}\big] \ \; \ 
\textrm{s.t.} \ \; \ \bm{S}\bm{\Delta h}=\bm{\Delta h_s}.  
\end{align}
where $\bm{\Delta h}$ is the residual vector to retrieve, $\bm{\Delta h_s}$ is the residual vector of measurements, and $\bm{L_d^{AW}}$ is an analogue Laplacian-based matrix to the one used in GSI, but obtained through \eqref{eq:3}.  


\subsection{Unscented Kalman Filter} The success of the Kalman filter to accurately estimate the state of a linear system led to the development of extensions to handle non-linear functions. The most notorious example is the Extended Kalman Filter (EKF), which exploits multivariate Taylor series expansions to linearize the model around the current estimate. This method has been successfully used in the past to estimate consumption and detect bursts in WDN \citep{Jung2015}.

The Unscented Kalman Filter (UKF) was designed to address the limitations of EKF, mainly related to the linearization precision. In EKF, only one point is considered to approximate a new linear function from the non-linear one, i.e., the mean of the Gaussian distribution which we assume that represents the form of our data. In UKF, a set of points known as Sigma Points are selected and mapped into the target Gaussian after being passed through the non-linear function. A process called Unscented Transformation helps recovering the approximated Gaussian after the application of the non-linear function. In comparison with EKF, UKF does not require the computation of the Jacobian, and the approximations are more accurate in the case of non-Gaussian inputs \citep{Julier1997}.

The UKF algorithm is well-established and has been implemented in several software platforms. Thus, we focus here on the adaptation of UKF to improve an initial estimation of the complete network state, represented by the nodal hydraulic heads. For more details about the UKF standard operation, 
see \cite{Wan2001}.

\section{Proposed Approach} \label{sec:approach}

Let us start by outlining the required input/output information, hyperparameters and the two key stages composing the UKF algorithm: prediction and data assimilation.


We discuss now the inputs and output of the estimation process.
First, $\bm h_0$ is the initial guess for the UKF, corresponding to the state of the network, i.e., the complete set of nodal heads. This first estimation may be retrieved from interpolation processes like GSI or AW-GSI.
Head measurements are stored in $\bm h_s$ from a set of $n_s$ pressure sensors. Considering that the number of network nodes is $n = |\mathcal{V}|$, normally $n >> n_s$.
The demand measurement vector $\bm{c_{a}}$ is constructed from the points where AMRs are installed. Again, if $n_{ca}$ AMRs are used, normally $n >> n_{ca}$.
Finally,
the output $\bm{h_{UKF}}$ of the UKF operation is a state estimation, obtained by fusing the information from the initial guess, the prediction function and the assimilation of the pressure and demand data.


Several configuration parameters must be settled before applying the UKF strategy.
Parameter $K$ corresponds to the total number of iterations of the UKF process. It must be configured through an analysis of the studied network (e.g. the selection of $K=50$ presented in Section \ref{sec:results}) or a convergence criteria, e.g., a tolerance value of the estimation change during a defined period.
Also, 
$\bm{Q}$ is a positive definite diagonal matrix denoting the covariance of the process noise. It accounts for the model approximations introduced by the prediction function.
Finally,
$\bm{R}$ is a positive definite diagonal matrix representing the covariance of the measurement noise, enabling us to express the level of confidence in the sensor data.

\subsection{Prediction} 

This process leverages a function that describes the state evolution from one time step to the next: 
    \begin{equation} \label{eq:4}
        \bm{h}^{[k+1]} = \textbf{f}(\bm{h}^{[k]}) = \alpha \bm{h}^{[k]} + (1-\alpha)\bm{\Phi}^{-1}\bm{\Omega} \bm{h}^{[k]}
    \end{equation}
    where $\bm{h}^{[k]}$ is the UKF state at iteration $[k]$\footnote{Note that index $k$ serves a dual purpose: indicating an arbitrary edge $\mathscr{e}_k$ and the $k\mbox{-}th$ iteration. To prevent confusion, $k$ only represents iteration number if it is encapsulated using brackets as $[k]$.}, $\bm{\Phi}$ and $\bm{\Omega}$ are respectively a degree matrix and a weighted adjacency matrix, and 
    $\alpha = \frac{n_{ca}}{n}$. In this way, \eqref{eq:4} yields a compromise solution between preserving or diffusing the current state, depending on the number of installed AMRs:
    if demand information is abundant, the state remains similar from one step to the next in terms of prediction, as most of the information is provided at the data assimilation step;
    else if demand information is scarce, the state information is diffused over the network by means of $\bm{\Phi}$ and $\bm{\Omega}$.

Besides, $\bm{\Omega}$ can be defined by the user, with $\bm{\Phi}$ obtained as $\phi_{ii} = \sum_{j=1}^{n} \omega_{ij}$. Possible selections for $\bm{\Omega}$ are the 0-1 adjacency matrix of $\mathcal{G}$, the GSI adjacency matrix $\bm{W}$ 
and the AW-GSI adjacency matrix $\bm{W^{AW}}$ (derived from \eqref{eq:3}).

\subsection{Data assimilation} 

This step uses a function that describes how the model states are related to sensor measurements. We have designed this function to account for the two available sources of measurements. 

First, the $n_s$ pressure sensors provide actual heads from the network, i.e., states of the UKF process. Thus, $\bm{y_h}^{[k]} = \bm{S}\bm{h}^{[k]}$
where $\bm{y_h}^{[k]}$ is the part of 
the measurement vector corresponding to the head data at iteration $[k]$.

Second, as $n_{ca}$ demand measurements are available, we can utilize the relation among nodal demands and hydraulic heads to derive the required function.  Starting with the Hazen-Williams equation, we have that
$m_{kj}(h_i-h_j) = \tau_{k}q_{k}^{1.852}$,
where $m_{kj}$ is analogue to \eqref{eq:8}, $\tau_{k} = \frac{1}{\sigma_{k}}$, and $q_{k}$ is the flow through the pipe represented by edge $\mathscr{e}_k$. 
This can be posed 
as:
\begin{equation}\label{eq:7}
\bm{M}\bm{h} = \bm{T}\bm{q}^{(1.852)},
\end{equation}
where  $\bm{q}^{(1.852)}=[q_1^{1.852} \;q_2^{1.852}\; ...\; q_{|\mathcal{E}|}^{1.852}]^T$ and $\bm{T}\in\mathbb{R}^{|\mathcal{E}|\times |\mathcal{E}|}$ is a diagonal matrix, whose $k\mbox{-}th$ diagonal value is $\tau_k$. 
Besides, the relation between nodal demands and flows can be expressed as:
\begin{equation}\label{eq:9}
\bm{c} = -\bm{M}^T\bm{q}
\end{equation}
with $\bm{c}\in\mathbb{R}^n$ denoting the vector of nodal demands.
Manipulating  \eqref{eq:7} and \eqref{eq:9} we obtain
$\bm{y_d}^{[k]} = -\bm{M}^T_a(\bm{T}^{-1}\bm{M}\bm{h}^{[k]})^{(0.54)}$
where $\bm{y_d}^{[k]}$ is the part of 
the measurement vector corresponding to the demand measurements at iteration $[k]$, and $\bm{M}_a$ selects only the columns of $\bm{M}$ that correspond to the nodes with AMRs.
Then, the complete measurement vector at iteration $[k]$ is
$\bm{y}^{[k]} = \textbf{g}(\bm{h}^{[k]}) = [\bm{y_h}^{[k]} \; \bm{y_d}^{[k]} ]^T$.
If the WDN measurements are stored in $\bm{y} = [\bm{h}_s \;\bm{c}_a]^T$, the measurement error for iteration $[k]$, used by the data assimilation process to update the current state, would be:
\begin{equation}\label{eq:12}
    \bm{e}^{[k]} = \bm{y} - \bm{y}^{[k]} = \bm{y} - \textbf{g}(\bm{h}^{[k]})
\end{equation}

\subsection{Dynamic-weighting prediction step}

In the past, the physics-based weight generation of AW-GSI led to an improvement in the state estimation performance. 
Therefore, these weighting mechanisms can be used to enhance the capabilities of the proposed UKF-based method, leading to 
dynamic update of the diffusion matrices in \eqref{eq:4}. Thus, this function is updated to:
\begin{equation} \label{eq:13}
        \bm{h}^{[k+1]} = \textbf{f}(\bm{h}^{[k]},\bm{\Omega}^{[k]},\bm{\Phi}^{[k]}) = \alpha \bm{h}^{[k]} + (1-\alpha)(\bm{\Phi}^{[k]})^{-1}\bm{\Omega}^{[k]} \bm{h}^{[k]}
    \end{equation}
where static matrices $\bm{\Omega}$ and $\bm{\Phi}$ have been substituted by their dynamic versions, i.e., $\bm{\Omega}^{[k]}$ and $\bm{\Phi}^{[k]}$, corresponding to iteration $[k]$. Then, the computation of $\bm{\Omega}^{[k]}$ would exploit the weighting generation of AW-GSI, via \eqref{eq:3}:
\begin{equation}\label{eq:14}
    \bm{\Omega}^{[k]} = 
    \begin{cases}         \bm{T}^{(-0.54)}\left(\bm{M}^{[k]}\bm{h}^{[k]}\right)^{(-0.46)}, & \mbox{mod}(k,K_u) = 0 \\
    \bm{\Omega}^{[k-1]}, & \mbox{otherwise.} 
    \end{cases}
\end{equation}
where $K_u$ denotes a user-defined iteration interval between consecutive weight updates (configurable analogously to $K$). A higher value for this parameter increases the number of iterations the UKF undergoes before reaching a final steady-state estimation, but contributes to greater stability before consecutive weight updates.

The initial guess of the state $\bm{h}_0$ and the weighted adjacency matrix $\bm{\Omega}^{[0]}$ can be retrieved from AW-GSI. Moreover, $\bm{\Phi}^{[k]}$ is obtained from $\bm{\Omega}^{[k]}$ as previously explained, and $\bm{M}^{[k]}$ is obtained through \eqref{eq:3} with $\tilde{\bm{h}} = \bm{h}^{[k]}$.



In order to complete the presented explanations, we present Algorithm \ref{alg:1} and Algorithm \ref{alg:2}, which respectively summarize the operational flow of the UKF-based approach with static prediction weights, henceforth denoted as UKF-GSI, and the upgraded version with dynamic prediction weights, denoted as UKF-AW-GSI. 
Please note that
steps 3 and 4 from both algorithms represent the UKF data assimilation and prediction steps respectively. Please see \cite{Wan2001} for additional details about the related equations and the role of $\bm{Q}$ and $\bm{R}$ in those steps.  
Also,
in Algorithm \ref{alg:2}, $\bm{W^{AW}}_0$ and $\bm{D^{AW}}_0$ correspond to the matrices used to derive $\bm{h}_0$ using AW-GSI.
Finally,
from a leak localization perspective, we must consider that most methods compare leak and leak-free scenarios, and hence the presented algorithms should be applied in both cases.
\begin{algorithm}[t]
\caption{UKF-GSI}
\label{alg:1}
\begin{algorithmic}[1]
\REQUIRE {$\bm{h}_0, \bm{y}, \bm{Q}, \bm{R}, K$}
\STATE Initialize: $\bm{h}^{[0]} = \bm{h}_0$
\FOR{$k$ = 0,\:...\:,\:$K-1$}
\STATE Correction \eqref{eq:12}: $\bm{h}^{[k]}\leftarrow\big(\bm{y} - \textbf{g}(\bm{h}^{[k]}), \bm{R}\big)$
\STATE Prediction \eqref{eq:4}: $\bm{h}^{[k+1]}\leftarrow\big( \textbf{f}(\bm{h}^{[k]}), \bm{Q}\big)$
\ENDFOR
\RETURN $\bm{h_{UKF}} = \bm{h^{[K]}}$
\end{algorithmic}
\end{algorithm}
\begin{algorithm}[t]
\caption{UKF-AW-GSI}
\label{alg:2}
\begin{algorithmic}[1]
\REQUIRE {$\bm{h}_0, \bm{y}, \bm{Q}, \bm{R}, \bm{W^{AW}}_0, \bm{D^{AW}}_0, K, K_u$}
\STATE Initialize: $\bm{h}^{[0]} = \bm{h}_0, \bm{\Omega}^{[0]} = \bm{W^{AW}}_0, \bm{\Phi}^{[0]} = \bm{D^{AW}}_0$
\FOR{$k$ = 0,\:...\:,\:$K-1$}
\STATE Correction \eqref{eq:12}: $\bm{h}^{[k]}\leftarrow\big(\bm{y} - \textbf{g}(\bm{h}^{[k]}), \bm{R}\big)$
\STATE Prediction \eqref{eq:13}: $\bm{h}^{[k+1]}\leftarrow\big(\textbf{f}(\bm{h}^{[k]}, \bm{\Omega}^{[k]}, \bm{\Phi}^{[k]}), \bm{Q}\big)$
\STATE AW update \eqref{eq:14}: $[\bm{\Omega}^{[k+1]},\bm{\Phi}^{[k+1]}]\leftarrow\big(\bm{h}^{[k]}, \bm{\Omega}^{[k]},K_{u}\big)$
\ENDFOR
\RETURN $\bm{h_{UKF}} = \bm{h^{[K]}}$
\end{algorithmic}
\end{algorithm}

\section{Case study} \label{sec:case_study}

\begin{figure}[t]
\begin{center}
\includegraphics[width=8.4cm]{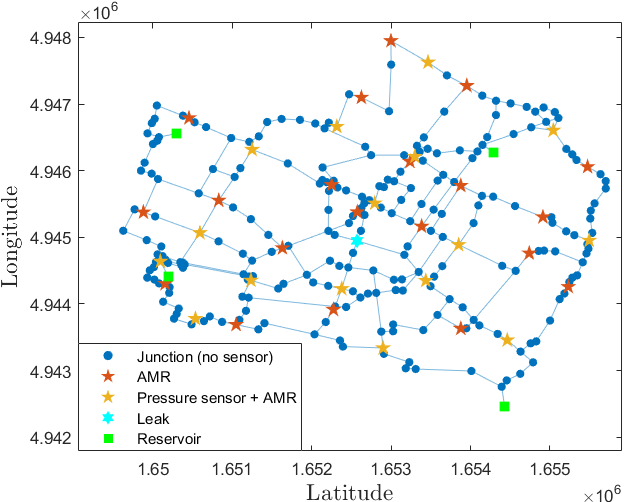}    
\caption{Associated graph to the Modena network.} 
\label{fig:1}
\end{center}
\end{figure}

The proposed methodology is tested by means of the Modena benchmark, which stands as a prominent, openly accessible case study within the management of WDNs \citep{Bragalli2012}. The network structure is represented in Fig.~\ref{fig:1}, whereas its main physical and hydraulic properties are introduced in Table \ref{table:1}. Note that the benchmark models a real-world network, whose size and demand correspond to a medium/large scale problem.
\begin{table}[t]
\begin{center}
\caption{Modena characteristics summary}\label{table:1}
\begin{tabular}{cc}
\toprule
\textbf{Property} & \textbf{Value} \\\midrule
Junctions & 268 \\
Pipes & 317\\
Reservoirs (water inlets) & 4\\
Total pipe length & 71.8 km\\
Total nodal demand & $\sim$400 $\ell$/s \\ \bottomrule
\end{tabular}
\end{center}
\end{table}
The pressure and demand sensors depicted in Fig.~\ref{fig:1} are assumed to be installed within the network. The pressure sensors include 4 metering devices at the network reservoirs and 16 at junctions. 
From the set of AMRs, 20 are located alongside the pressure sensors, whereas the other 20 are placed in additional locations. All these placements are obtained through a fully data-driven sensor placement technique, presented in \cite{RomeroBen2022c}. 

\subsubsection{Generation of evaluation data}

The presented benchmark is implemented in 
EPANET 2.0 \citep{Epanet2000} to obtain hydraulic data from leak and nominal scenarios, 
enabling the assessment of the method's performance. 

The conducted simulations span a 24-hour period, during which the nodal demands evolve with a pattern that changes every hour. 
A nodal extra demand of 4.5 $\ell$/s emulates the leak effects. We consider this leak size to be adequate, regarding that it only accounts up for a $\sim$1.1\% of the average total inflow, as well as recent studies dealing with the same benchmark consider similar or even larger leak sizes \citep{Alves2022}. The leak size selection is also justified by the 
included sources of uncertainty as follows.
First,
the system relies on the accuracy of the measured hydraulic values, i.e., pressure and consumption, within a margin of $\pm1$ cm and $\pm0.01$ $\ell$/s respectively.
Second,
random uncertainty has been added to the pipes' diameter and roughness, which are typically difficult to measure in WDNs. An uncertainty level of 1\% with respect to the noise-free values is included.
Furthermore, daily demand patterns are also affected by a 1\% of uniformly random uncertainty, including additional variability to the actual consumption, and therefore the produced nodal pressures.


\section{Results AND Discussion\protect\footnotemark}\label{sec:results}

\footnotetext{Code and data: \url{https://github.com/luisromeroben/UKF-AW-GSI}}

The efficacy of AW-GSI has previously been validated for the presented case study, producing satisfactory state estimations that led to successful leak localization. However, performance issues emerged in certain leak scenarios, leading to deficient estimation and/or localization outcomes.

We showcase here state estimation and leak localization results for a challenging leak case, illustrating the behavior of the newly proposed methods in such a scenario that led to degraded solutions through AW-GSI. Specifically, we consider a leak at node 88, which can be seen in Fig.~\ref{fig:1} (labeled as "Leak") to be positioned away from pressure or demand sensors, which is challenging for the compared methodologies. 
The configured parameters for the UKF-based approaches are listed in Table \ref{table:2} where $\bm{I}_n$ is the identity matrix of size $n$. The amount of sensors was selected as part of the problem definition, as previously explained. The rest of parameters was empirically configured. 
\begin{table}[t]
\begin{center}
\caption{UKF-based method parameters}\label{table:2}
\begin{tabular}{ccccccc}
\toprule
\textbf{Parameter} & $n_s$ & $n_{ca}$ & $K$ & $K_u$ & $\bm{Q}$ & $\bm{R}$   \\ 
\midrule
\textbf{Value} & 20 & 40 & 50  & 5  & $\bm{I}_n$ & $1e^{-4}\bm{I}_n$ \\
\bottomrule
\end{tabular}
\end{center}
\end{table}

\subsection{State estimation}

The state estimation performance is studied through the degree of dissimilarity between the actual and reconstructed hydraulic head vectors, i.e., $\bm{h}$ and $\bm{h_{UKF}}$. To this end, we compute the root-squared-mean error as $RMSE(\bm{h}, \bm{h_{UKF}}) = \sqrt{\frac{1}{n} (\bm{h}-\bm{h_{UKF}})^T(\bm{h}-\bm{h_{UKF}})}$. 



\begin{figure}[t]
\begin{center}
\includegraphics[width=8.4cm]{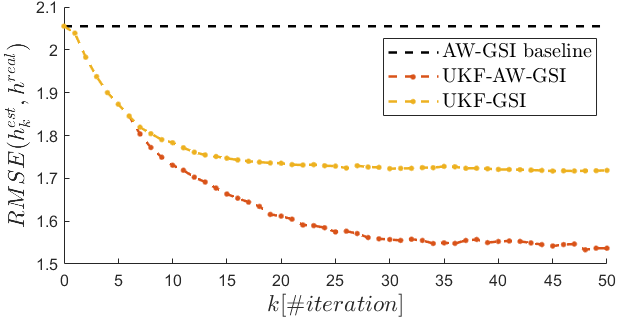}    
\caption{State $RMSE$ evolution comparison.} 
\label{fig:2}
\end{center}
\end{figure}

Fig.~\ref{fig:2} shows the $RMSE$ evolution through the UKF iterations for a challenging time instant, which yielded the worst AW-GSI estimation performance among all the available ones, for different estimation methodologies and configuration settings. In this way, the improvement of the new methods through the iterations can be observed. We use AW-GSI as a baseline (note that it does not iterate, so its $RMSE$ is depicted as a horizontal line) for comparison with UKF-GSI and UKF-AW-GSI. 
Both methods improve the AW-GSI estimation (used as $\bm{h}_0$), demonstrating the suitability of the devised UKF-based scheme. Moreover, UKF-AW-GSI performs better than UKF-GSI, specifically yielding a significant refinement when $\mbox{mod}(k,K_u)=0$, enabling the AW update step in \eqref{eq:14}. Specifically, this implies a $RMSE$ reduction of 16.38\% for UKF-GSI and 25.22\% for UKF-AW-GSI with respect to AW-GSI.

In order to underscore the importance of employing a suitable initial guess, we perform the same experiment with UKF-AW-GSI, first, with $\bm{h}_0 = \bm{0}_n$, 
and second,
with a vector obtained as $\bm{h}_0 = \mu(\bm{h^{AW}}) + \sigma(\bm{h^{AW}})\bm{x}$, with $\mu(\bm{h^{AW}})$ and $\sigma(\bm{h^{AW}})$ being the mean and standard deviation of the state vector retrieved from AW-GSI, i.e., $\bm{h^{AW}}$; and $\bm{x}\in\mathbb{R}^{n}$ is a random vector in $[-1, 1]$. The simulations show an initial RMSE of 51.11m and 9.38m respectively ($k=0$) which is reduced at $k=50$ to 11.42m and 2.62m respectively.
Both results do not even reach the baseline RMSE from Fig.~\ref{fig:2}.

Finally, let us study the performance from a general perspective, extending the previous analysis to a wide set of time instants, thus achieving results in different network conditions. Specifically, we select one hour out of every two, leading to a total of 12 considered time instants, used to ultimately compute the RMSE vector $\bm{r}$. Table \ref{table:3} shows the performance $RMSE$-based results for AW-GSI, UKF-GSI and UKF-AW-GSI through various statistics. 
\begin{table}[t]
\begin{center}
\caption{$RMSE$ comparison summary (m)}\label{table:3}
\begin{tabular}{cccc}
\toprule
\textbf{Method} & \textbf{$\mu(\bm r)\pm\sigma(\bm r)$} & \textbf{$\mbox{max}(\bm r)$} & \textbf{$\mbox{min}(\bm r)$}   \\ 
\toprule
AW-GSI & 1.26 $\pm$ 0.59 & 2.06 & 0.35 \\
\midrule
UKF-GSI & 1.06 $\pm$ 0.49 & 1.72 & 0.31 \\
\midrule
UKF-AW-GSI & 0.95 $\pm$ 0.44 & 1.54 & 0.28  \\
\bottomrule
\end{tabular}
\end{center}
\end{table}


\subsection{Leak localization}

The presented state estimation methodology is proposed in the context of fault isolation. Thus, analysing the localization result from a leak localization method using the reconstructed states can illustrate the adequacy of the estimation approach. To this end, Leak Candidate Selection Method (LCSM) has been chosen due to its data-driven operation, as well as its satisfactory performance with GSI-based estimation methods \citep{RomeroBen2022b}. It compares leak and leak-free (with similar boundary conditions) states, deriving a distance-based metric that serves as indicator of the leak likelihood. 
\begin{figure}[t]
\begin{center}
\includegraphics[width=8.4cm]{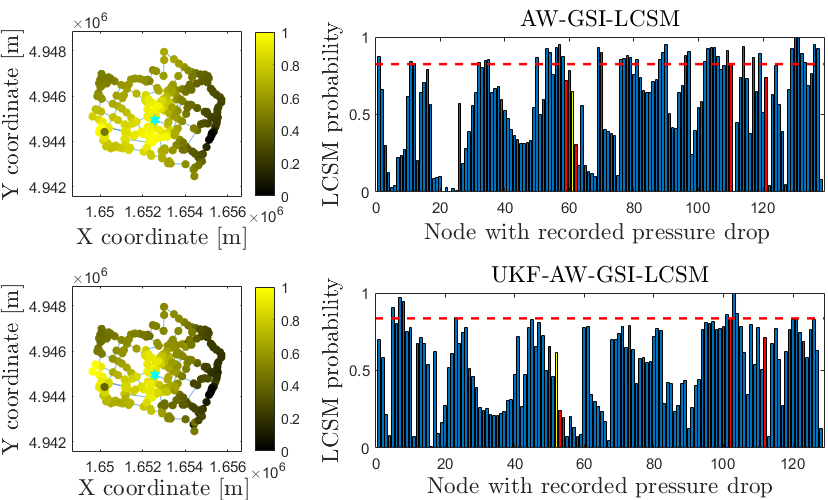}    
\caption{Localization performance comparison between AW-GSI-LCSM and UKF-AW-GSI-LCSM for the selected leak scenario.} 
\label{fig:3}
\end{center}
\end{figure}

The localization results for the studied leak scenario are presented in Fig.~\ref{fig:3}, considering the same hydraulic data as in Table \ref{table:3}. First, a colour map over the network graph is depicted for AW-GSI-LCSM and UKF-AW-GSI-LCSM in the left part of the figure. This graph provides a colour to each node depending on their associated value of the LCSM distance metric: the more yellow the node, the higher the probability of being the leak location. Then, a bar graph is presented for each method in the right part of the figure. The x-axis denotes the node index, but taking into account that only nodes with a positive LCSM metric are represented, because this implies the existence of a pressure drop. On the y-axis, we represent the LCSM metric. The yellow bar indicates the leak node (88), whereas the red bars represent its neighbours. 

A preliminary analysis using the colour maps highlights the challenging nature of this leak scenario. Although the area around the leak has a medium-to-high probability in both methodologies, the most probable locations are rather far from the correct node (left part of the WDN). However, the in-depth analysis shown through the bar plots manifest the improvements of UKF-AW-GSI-LCSM. The red horizontal dashed line indicates the maximum value of the LCSM likelihood metric among the neighbours of the leak node. Thus, in the case of AW-GSI-LCSM, there are 40 nodes with a higher LCSM metric than the best candidate among the set composed by the leak and its first-degree neighbours, whereas in the case of UKF-AW-GSI-LCSM, there are only 7 (corresponding to the aforementioned left part of the network). 

This shows the promising performance of the UKF-based scheme, whose better state estimation helps to improve the accuracy of the localization process. Note that UKF-AW-GSI inherits the estimation of AW-GSI through the initial guess, so the UKF-based approach demonstrates itself capable of enhancing the estimation to the point of reducing the importance given to incorrectly indicated areas by AW-GSI-LCSM during the localization step.


\section{Conclusions} \label{sec:conclusions}

This article has presented a nodal hydraulic head estimation methodology for WDN based on an UKF scheme with application to data-driven leak localization. This technique leverages the state reconstruction capabilities of the UKF, customizing the prediction and data assimilation functions through information about the physics behind network dynamics, its structure, and the available pressure and demand measurements. An improved version is devised by including a dynamic weighting approach that updates the weights included in the prediction step. The performance of the strategy is tested using the Modena benchmark. The UKF-based schemes provided improvements over AW-GSI in terms of estimation error and leak localization, showing the adequateness of the methods.  

Future work will include a deep analysis of the effects of the parametrization on the performance, as well as the degradation caused by uncertainty, designing new mechanisms to reduce the effect of noise. Improvements in the operational flow of the method will be researched, trying to improve the head estimation. Additionally, the estimation of other related hydraulic variables such as demand or flow will be considered.



\bibliography{ifacconf}             

\end{document}